\documentclass[twocolumn,prl,amsmath,amssymb]{revtex4}
\usepackage{graphicx}
\begin{document}

\title{Helicity Entanglement of Moving Bodies}

\date{\today}
\author{Shuxin Shao}
\affiliation{Department of Physics, Beijing Normal University,
Beijing, 100875, China}
\author{Song He}
 \affiliation{School of Physics, Peking University, Beijing, 100871,
China}
\author{Hongbao Zhang}
\email{hzhang@perimeterinstitute.ca}
    \affiliation{Perimeter Institute for Theoretical Physics, Waterloo, Ontario, N2L 2Y5, Canada\\
    Department of Applied Mathematics, University of Waterloo, Waterloo, Ontario, N2L 3G1, Canada}

\begin{abstract}
We investigate the Lorentz transformation of the reduced helicity
density matrix for a pair of massive spin $\frac{1}{2}$ particles.
The corresponding Wootters concurrence shows no invariant meaning,
which implies that we can generate helicity entanglement simply by the transformation from one reference frame to
another. The difference between the helicity and
spin case is also discussed.
\end{abstract}


\maketitle As a central concept, quantum entanglement is the
major resource in quantum information science such as quantum
teleportation and quantum computation\cite{Bouwmeester}. Therefore
in quantum information science, one of the
most critical issues is how to generate quantum entanglement. In
general, quantum entangled states are prepared through some kinds of
dynamical processes. However, it is recently shown in the domain of relativistic quantum information that quantum
entanglement has somewhat observer dependent property\cite{Peres1}. Speaking specifically, for a
single free massive spin $\frac{1}{2}$ particle, the spin entropy,
measurement of entanglement between spin and momentum degrees of
freedom, has no invariant meaning under the transformation of inertial reference frames\cite{Peres2}. In
particular, even though the initial state is a direct product of a
function of momentum and a function of spin, the Lorentz transformed
state is not a direct product, which means that spin and momentum
appear to be entangled. Nonetheless, this is not the familiar type
of entanglement available for quantum communication, because both
degrees of freedom belong to the same particle, rather than to the
distinct subsystems that could be widely separated. Based on this
observation, Gingrich and Adami have investigated the Lorentz
transformation properties of entanglement between a pair of massive
spin $\frac{1}{2}$ particles\cite{Gingrich}. As a result, while the
entanglement of the two particles' entire wave function, i.e., both
momentum and spin included, is Lorentz invariant, their spin or momentum entanglement
 may change due to Lorentz
transformation. Thus in some sense only by transforming one
reference frame to another does relativity provide a brand new
alternative road to create quantum entanglement, which could be used
for entanglement manipulation.

As inferred above, in relativistic quantum information, previous investigations focus primarily on
the spin relevant entanglement\cite{Alsing,Pachos,Caban,Lamata1,Jordan1,Lamata2,Jordan2}. But it
is the helicity rather than spin that is more often under both
theoretical consideration and experimental detection in high energy
physics, because the helicity has an advantage in providing a smooth
transition to the massless case. Although both the helicity states
and spin states can constitute the basis of Hilbert space of one
particle, they differ in the way of unitary transformation under the
action of Lorentz group\cite{Weinberg}. As analyzed most recently
for a single free massive spin $\frac{1}{2}$ particle, the
entanglement properties for helicity differ remarkably from those
for spin after we trace out the momentum degree of freedom under
Lorentz transformation\cite{He}. Nevertheless, obviously it is more intriguing and
significant to investigate the Lorentz transformation properties for
helicity entanglement between a pair of massive spin $\frac{1}{2}$
particles, which is just what we shall report in this paper.

Given a field with mass $m$ and spin $\frac{1}{2}$, we can also
construct the helicity states $|p;\lambda\rangle$ as a complete
orthonormal basis for Hilbert space of one particle. Associated with
a Lorentz transformation $\Lambda$ the unitary operator $U(\Lambda)$
acting on these helicity states gives\cite{Weinberg}
\begin{eqnarray}
&&U(\Lambda)|p;\lambda\rangle \nonumber\\
&&=\sqrt{\frac{(\Lambda
p)^0}{p^0}}D_{\lambda'\lambda}[R^{-1}(\Lambda p)L^{-1}(\Lambda
p)\Lambda L(p)R(p)]|\Lambda p;\lambda'\rangle \nonumber\\
&&=\sqrt{\frac{(\Lambda
p)^0}{p^0}}D_{\lambda'\lambda}[B^{-1}(\Lambda p)R^{-1}(\Lambda
p)\Lambda R(p)B(p)]|\Lambda p;\lambda'\rangle \nonumber\\
&&=\sqrt{\frac{(\Lambda
p)^0}{p^0}}D_{\lambda'\lambda}[Z(\Lambda,p)]|\Lambda
p;\lambda'\rangle. \label{LT}
\end{eqnarray}
Here $R(p)$ is the rotation that carries the $z$ axis into the
direction $\mathbf{p}$, $B(p)$ is the boost from rest to the
momentum $|\mathbf{p}|$ in the $z$ direction, and $L(p)$ is the pure
boost from rest to the momentum $\mathbf{p}$. Obviously,
$L^{-1}(\Lambda p)\Lambda L(p)$ is just Wigner rotation, usually
denoted by $W(\Lambda,p)$. In addition, $D$ is the spin
$\frac{1}{2}$ irreducible unitary representation of Lorentz group.
Note that these helicity states differ in way of unitary
transformation from spin states under the action of Lorentz group,
since under Lorentz transformations spin states change according to
Wigner rotation, which is related to $Z(\Lambda,p)$ as
$Z(\Lambda,p)=R^{-1}(\Lambda p)W(\Lambda,p)R(p)$.

Thus a pure state for two massive spin $\frac{1}{2}$ particles can
be expanded in terms of the helicity states as
\begin{equation}
|\Psi\rangle=\int
d\mathbf{p}d\mathbf{q}\sum_{\lambda\sigma}g_{\lambda\sigma}(\mathbf{p},\mathbf{q})|p;\lambda\rangle|q;\sigma\rangle,
\end{equation}
with the normalized condition
\begin{equation}
\int
d\mathbf{p}d\mathbf{q}\sum_{\lambda\sigma}|g_{\lambda\sigma}(\mathbf{p},\mathbf{q})|^2=1.
\end{equation}
Then the corresponding reduced helicity density matrix can be
obtained by tracing out the momentum degrees of freedom, i.e.,
\begin{eqnarray}
\rho&=&\mathrm{Tr}_{\mathbf{p},\mathbf{q}}[|\Psi\rangle\langle\Psi|]=\int
d\mathbf{p}d\mathbf{q}\langle\mathbf{p},\mathbf{q}|\Psi\rangle\langle\Psi|\mathbf{p},\mathbf{q}\rangle\nonumber\\
&=&\int
d\mathbf{p}d\mathbf{q}\sum_{\lambda\tilde{\lambda}\sigma\tilde{\sigma}}
[g_{\lambda\sigma}(\mathbf{p},\mathbf{q})g^*_{\tilde{\lambda}\tilde{\sigma}}(\mathbf{p},\mathbf{q})|\lambda,\sigma\rangle\langle\tilde{\lambda},\tilde{\sigma}|],\nonumber\\
\end{eqnarray}
where we have employed the orthonormal relation for the helicity
states.

The entanglement between the helicity degrees of freedom can be
quantified by calculating Wootters concurrence\cite{Wootters}
\begin{equation}
C(\rho)=\mathrm{Max}\{k_1-k_2-k_3-k_4,0\},
\end{equation}
where $\{k_1\geq k_2\geq k_3\geq k_4\}$ are the square roots of the
eigenvalues of the matrix $\rho\tilde{\rho}$ with the time reversed
matrix
\begin{equation}
\tilde{\rho}=\big[\left(
               \begin{array}{cc}
                 0 & -i\\
                 i & 0 \\
               \end{array}
             \right)\otimes\left(
               \begin{array}{cc}
                 0 & -i\\
                 i & 0 \\
               \end{array}
             \right)\big]\rho^*\big[\left(
               \begin{array}{cc}
                 0 & -i\\
                 i & 0 \\
               \end{array}
             \right)\otimes\left(
               \begin{array}{cc}
                 0 & -i\\
                 i & 0 \\
               \end{array}
             \right)\big].\\
\end{equation}

Note that under a Lorentz transformation $\Lambda$ the two particle
state changes as follows
\begin{eqnarray}
&&|\Psi'\rangle=U(\Lambda)|\Psi\rangle=\int d\mathbf{p}
d\mathbf{q}\sum_{\lambda\lambda'\sigma\sigma'}\sqrt{\frac{(\Lambda
p)^0}{p^0}}\sqrt{\frac{(\Lambda q)^0}{q^0}}\nonumber\\
&&g_{\lambda\sigma}(\mathbf{p},\mathbf{q})D_{\lambda'\lambda}[Z(\Lambda,p)]D_{\sigma'\sigma}[Z(\Lambda,q)]|\Lambda
p;\lambda'\rangle|\Lambda q;\sigma'\rangle.\nonumber\\
\end{eqnarray}
Accordingly one can obtain the transformed reduced helicity density
matrix as
\begin{eqnarray}
&&\rho'=\int
d\mathbf{p}d\mathbf{q}\sum_{\lambda\lambda'\sigma\sigma'}^{\tilde{\lambda}\tilde{\lambda}'\tilde{\sigma}\tilde{\sigma}'}D_{\lambda'\lambda}[Z(\Lambda,p)]D_{\sigma'\sigma}[Z(\Lambda,q)]g_{\lambda\sigma}(\mathbf{p},\mathbf{q})\nonumber\\
&&g^*_{\tilde{\lambda}\tilde{\sigma}}(\mathbf{p},\mathbf{q})D^\dagger_{\tilde{\lambda}\tilde{\lambda}'}[Z(\Lambda,p)]D^\dagger_{\tilde{\sigma}\tilde{\sigma}'}[Z(\Lambda,q)]|\lambda',\sigma'\rangle\langle\tilde{\lambda}',\tilde{\sigma}'|.\nonumber\\
\end{eqnarray}
By Eq.(\ref{LT}), $D[Z(\Lambda,p)]$ is always an identity matrix if
$\Lambda$ is a purely spacial rotation transformation, which means
that the reduced helicity matrix is completely the same among those
inertial observers without relative motion but with different
identification of spacial direction. This property stems essentially 
from the fact that the helicity
$\frac{(\mathbf{p}\cdot\mathbf{S})}{|\mathbf{p}|}$ remains invariant
under a purely spacial rotation transformation, where $S$ denotes
the spin\cite{spin}. Note that any Lorentz transformation can always
be decomposed into the product of a pure boost and a pure rotation.
We thus shall concentrate on what happens to the reduced helicity
matrix and the corresponding Wootters concurrence when $\Lambda$ is
a pure boost transformation. Furthermore, the pure boost transformations are similarly equivalent with one
another by rotations, which means the reduced helicity density
matrix only depends on the speed of relative motion
between the inertial observers. Therefore we shall only need to consider the pure
boost transformations along the $z$ axis.

In the special case mentioned above, set
\begin{equation}
\Lambda=\left(
          \begin{array}{cccc}
            \cosh\eta & 0 & 0 & \sinh\eta \\
            0 & 1 & 0 & 0 \\
            0 & 0 & 1 & 0 \\
            \sinh\eta & 0 & 0 & \cosh\eta \\
          \end{array}
        \right),\eta\leq 0,
\end{equation}
and
\begin{equation}
 p=m[\cosh\tau,\sinh\tau(\sin\theta\cos\phi,\sin\theta\sin\phi,\cos\theta)],\tau\geq
 0,
\end{equation}
then employing Eq.(\ref{LT}), we obtain
\begin{eqnarray}
&&D[Z(\Lambda,p)]=\left(
                    \begin{array}{cc}
                      e^{-\frac{\alpha}{2}} & 0 \\
                      0 & e^\frac{\alpha}{2} \\
                    \end{array}
                  \right)\times\nonumber\\
&&\left(
                                                      \begin{array}{cc}
                                                       \cos\frac{\beta}{2} & \sin\frac{\beta}{2} \\
                                                        -\sin\frac{\beta}{2} & \cos\frac{\beta}{2} \\
                                                      \end{array}
                                                    \right)\left(
                 \begin{array}{cc}
                   e^\frac{\eta}{2} & 0 \\
                   0 & e^{-\frac{\eta}{2}} \\
                 \end{array}
               \right)\times
\nonumber\\
   &&\left(
                                                      \begin{array}{cc}
                                                       \cos\frac{\theta}{2} & -\sin\frac{\theta}{2} \\
                                                        \sin\frac{\theta}{2} & \cos\frac{\theta}{2} \\
                                                      \end{array}
                                                    \right)\left(
                                                             \begin{array}{cc}
                                                               e^\frac{\tau}{2} & 0 \\
                                                               0 & e^{-\frac{\tau}{2}} \\
                                                             \end{array}
                                                           \right),\label{matrix}
\end{eqnarray}
where $\alpha$ and $\beta$ satisfy
\begin{equation}
\cosh\alpha=\cosh\eta\cosh\tau+\sinh\eta\sinh\tau\cos\theta,\alpha\geq
0,
\end{equation}
and
\begin{equation}
\cos\beta=\frac{\sinh\eta\cosh\tau+\cosh\eta\sinh\tau\cos\theta}{\sqrt{\sinh^2\tau\sin^2\theta+(\sinh\eta\cosh\tau+\cosh\eta\sinh\tau\cos\theta)^2}}
\end{equation}
with $\pi\geq\beta\geq 0$, respectively.

\begin{figure}[t]
\begin{center}
\includegraphics[clip=truth,width=1.0\columnwidth]{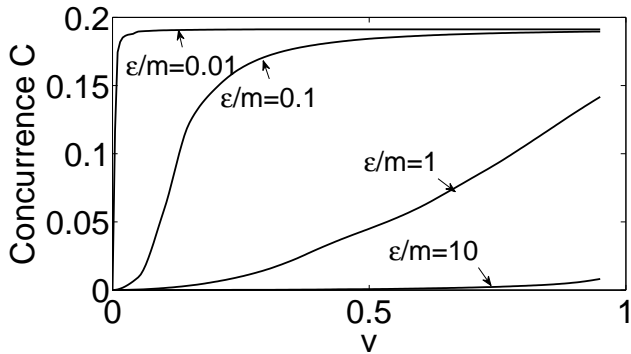}
\caption{The Wootters concurrence $C$ as a function of the speed
$v=-\tanh\eta$ of the inertial observers with respect to the original
inertial reference frame.}\label{helicity}
\end{center}
\end{figure}

For simplicity but without loss of generalization, choose in
particular the initial normalized state with the nonvanishing
component as follows
\begin{equation}
g_{\frac{1}{2}\frac{1}{2}}(\mathbf{p},\mathbf{q})=f(\mathbf{p})f(\mathbf{q}),
\end{equation}
where the momentum distribution function $f$ is taken as a Gaussian.
i.e.,
\begin{equation}
f(\mathbf{p})=\pi^{-\frac{3}{4}}\varepsilon^{-\frac{3}{2}}e^{-\frac{\mathbf{p}^2}{2\varepsilon^2}}
\end{equation}
with the distribution width $\varepsilon$. It is obvious that this
initial state is an unentangled helicity state, and its Wootters
concurrence vanishes. However, with respect to the moving inertial
observers, the corresponding concurrence is greater than zero in
general, which is carried out by numerical calculations and is
specifically illustrated in Fig.\ref{helicity}.

As shown in Fig.\ref{helicity}, with the speed of the
inertial observers increasing, the variation behavior of the corresponding
concurrence is dependent upon the width-mass ratio
$\frac{\varepsilon}{m}$. In particular, for the large width-mass
ratio limiting case, the resultant concurrence varies pretty
slightly and remains nearly invariant. On the other hand, for the
limiting case of sharp momenta which corresponds to the small
width-mass ratio, the concurrence blows up from zero and rapidly
saturates. Speaking specifically, it arrives at a constant value at
small velocities of the observers, and then remains nearly invariant
regardless of the increase of speed of the observers. So, if we restrict
ourselves to helicity measurements in this case, an observer in the
original inertial reference frame cannot use quantum entanglement as
a resource while the moving observers can, simply with small
velocities. This is remarkably different from the spin case, since the concurrence for spin always remains invariant in the limit of sharp momenta\cite{Gingrich,Alsing,Pachos}.

The crucial physics underlying this behavior lies in the following
fact: With the smaller width-mass ratio $\frac{\varepsilon}{m}$,
i.e., sharper momentum distribution, the helicity of the prepared
particles becomes more sensitive to Lorentz boost, due to the
concentration of its momentum in a smaller neighborhood around zero.
To put it another way, for the case of smaller width-mass ratio, the
smaller speed of observers is needed to make the flip of helicity
from right to left saturate such that the corresponding concurrence
approaches the saturated value.

In summary, the transformation of the reduced helicity density
matrix under Lorentz group is investigated for a pair of massive
spin $\frac{1}{2}$ particles. Especially, we have calculated the
corresponding Wootters concurrence for an initial unentangled
helicity state by numerical computation. Our results show that the
corresponding concurrence is not a Lorentz invariant scalar, which
thus opens a new alternative way to generate quantum entanglement
between helicity degrees of freedom. In addition, as the speed of
the inertial observers increases, the specific variation of concurrence
for the helicity demonstrates a distinct behavior from the spin case, which essentially originates from the fact that the
helicity states differ significantly from the spin states in the
transformation property under Lorentz boost, as pointed out in the
beginning.

HZ is grateful to helpful suggestions from Achim Kempf. He would
also like to give much thanks to Robin Blume-Kohout for valuable
communication. Work by SH was supported by NSFC(nos.10235040 and
10421003). HZ was supported in part by the Government of China
through CSC(no.2007102530). This research was supported by Perimeter
Institute for Theoretical Physics. Research at Perimeter Institute
is supported by the Government of Canada through IC and by the
Province of Ontario through MRI.


\begin{thebibliography}{99}
\bibitem{spin}Note that the reduced spin density matrix changes as $\rho'=\big[D(\Lambda)\otimes D(\Lambda)\big]\rho
\big[D^\dagger(\Lambda)\otimes D^\dagger(\Lambda)\big]$ for a purely
spacial rotation transformation $\Lambda$. However, the
corresponding concurrence remains invariant under such a rotation.
\end{thebibliography}
\end{document}